\documentclass[10pt]{iopart}

\usepackage{iopams}
\usepackage{graphicx} 
\usepackage{dcolumn}
\usepackage{upgreek }
\usepackage{times}
\usepackage{textcomp}
\usepackage{ wasysym }
\usepackage{titlesec}
\usepackage{float}
\usepackage{color,xcolor}
\usepackage{graphicx,times} 
\usepackage{cuted} \stripsep -3pt plus 3pt minus 2pt
\usepackage{epstopdf}
\usepackage{graphicx}
\usepackage{subfigure}
\usepackage{makecell}
\usepackage{multirow}
\renewcommand{\figurename}{Figure}
\usepackage[colorlinks,linkcolor=blue, urlcolor=blue, anchorcolor=blue, citecolor=blue]{hyperref}
\usepackage{longtable,booktabs}
\usepackage{fouriernc}
\usepackage{epsfig,graphics,graphicx}
\usepackage{bm}
\usepackage{color}
\usepackage{flushend}
\usepackage{ulem}
\usepackage[sort&compress,square,numbers]{natbib}

\begin{document}
\title [Minimum Number of Parameters for Grad-Shafranov Solutions] {What Is the Minimum Number of Parameters Required to Represent Solutions of the Grad-Shafranov Equation?}

\author{
Huasheng Xie$^{1,2,*}$ and Yueyan Li$^{3}$
}

\address{

$^1$ Hebei Key Laboratory of Compact Fusion, Langfang 065001, People's Republic of China

$^2$ ENN Science and Technology Development Co., Ltd., Langfang 065001, People's Republic of China

$^3$ School of Science, Tianjin University of Technology and Education, Tianjin 300072, People's Republic of China

}
\eads{\mailto{huashengxie@gmail.com, xiehuasheng@enn.cn}}

\begin{indented}
\item[\today]
\end{indented}

\begin{abstract}
Fast and accurate solutions of the Grad--Shafranov (GS) equation are essential for equilibrium analysis, integrated modeling, and surrogate model construction in magnetic confinement fusion. In this work, we address a fundamental question: what is the minimum number of free parameters required to accurately represent numerical solutions of the GS equation under fixed-boundary conditions?
We demonstrate that, for most practical applications, GS equilibria can be represented using only 2--5 free parameters while maintaining relative errors below 5\%. For higher-accuracy requirements, we introduce a unified spectral representation based on the Miller extended harmonic (MXH) expansion in the poloidal direction combined with shifted Chebyshev (Cheb) polynomials in the radial direction. This MXH--Cheb basis exhibits rapid convergence for two-dimensional GS equilibria.
For configurations where three geometric moments (shift, elongation, and triangularity) are specified at the last closed flux surface (LCFS), relative errors on the order of $10^{-2}$--$10^{-3}$ can be achieved using as few as 13--20 parameters. In more general cases, including up--down asymmetric equilibria, X-point configurations, and stiff pressure and current profiles (e.g., H-mode pedestals), accuracies beyond this level can be obtained with fewer than 100 parameters. The resulting equilibrium configurations and profile functions are fully analytical, with smooth derivatives of all orders.
These results provide a systematic foundation for developing high-fidelity, ultra-fast GS solvers and enable efficient reduced-order and AI-based surrogate modeling of tokamak equilibria.
\end{abstract}

\maketitle
\ioptwocol

\section{Introduction}\label{sec:intro}

In the study of magnetized controlled fusion plasmas, the determination of equilibrium states is the foundational step for virtually all subsequent analyses, including stability assessment, transport simulation, and heating optimization. For axisymmetric configurations such as tokamaks, this problem reduces to solving the Grad--Shafranov (GS) equation from magnetohydrodynamic (MHD) equilibrium theory. In the cylindrical coordinate system $(R, \phi, Z)$, the GS equation is expressed as \cite{Freidberg1987, Jardin2010}:
\begin{equation}
\label{eq:GS_op}
\Delta^* \psi \equiv R \frac{\partial}{\partial R} \left( \frac{1}{R} \frac{\partial \psi}{\partial R} \right) + \frac{\partial^2 \psi}{\partial Z^2} = -\mu_0 R J_\phi,
\end{equation}
where $\psi$ is the poloidal magnetic flux per radian, and $\mu_0$ is the permeability of free space. The toroidal plasma current density $J_\phi$ is given by:
\begin{equation}
\label{eq:J_phi}
J_\phi = R P'(\psi) + \frac{F(\psi) F'(\psi)}{\mu_0 R},
\end{equation}
with the pressure profile $P(\psi)$ and the poloidal current function $F(\psi) = R B_\phi$. Here, the prime denotes differentiation with respect to $\psi$. The magnetic field components are related to the flux by:
\begin{equation}
B_R = -\frac{1}{R}\frac{\partial \psi}{\partial Z}, \quad B_Z = \frac{1}{R}\frac{\partial \psi}{\partial R}, \quad B_\phi = \frac{F(\psi)}{R},
\end{equation}
and the poloidal flux is typically defined as $\psi = \frac{1}{2\pi} \int B_Z R dR$.

Numerical solution of equation (\ref{eq:GS_op}) is a standard procedure in fusion research. However, a significant dichotomy exists in current methods regarding the \textit{representation} of the solution $\psi(R,Z)$. In this work, we limit our discussion to the fixed-boundary equilibrium problem, where the geometry of the last closed flux surface (LCFS) and the profile functions $P(\psi)$ and $F(\psi)$ are specified.

High-fidelity solvers typically rely on dense spatial discretizations or high-order spectral expansions. For instance, codes like CHEASE \cite{Lutjens1996} employ finite element methods (e.g., bicubic Hermite polynomials) on discrete grids, while ECOM \cite{Lee2015} combines conformal mapping with Fourier and integral equation methods. While mathematically robust, these approaches suffer from high dimensionality---often requiring $10^3 \sim 10^4$ degrees of freedom to accurately resolve steep gradients or boundary features---resulting in computational costs ranging from seconds to minutes. This computational burden is prohibitive for large-scale parameter scans or system codes \cite{Haney1992,Chen2026}. Alternatively, modern 3D equilibrium codes such as DESC \cite{Dudt2020} utilize global basis functions (e.g., Fourier--Zernike). However, when standard Fourier series are employed to describe the poloidal geometry, they become inefficient for modern tokamak shapes. The strong shaping (e.g., high elongation, D-shape) and the presence of X-points introduce sharp geometric features that are difficult to approximate with smooth sinusoids. Consequently, such expansions often suffer from slow spectral convergence and the Gibbs phenomenon, necessitating a large number of harmonic coefficients to suppress non-physical oscillations.

At the opposite end of the spectrum, reduced-order models, such as the variational methods by Haney et al. \cite{Haney1988, Haney1995} and Ludwig \cite{Ludwig1995}, utilize a minimal set of geometric parameters (typically 2--7 parameters, such as the Shafranov shift $\Delta$, elongation $\kappa$, and triangularity $\delta$). These models are extremely fast but are constrained by rigid geometric assumptions, making them insufficient for accurately representing complex, highly asymmetric configurations. Nevertheless, these variational parameter methods perform surprisingly well globally, often surpassing analytical large-aspect-ratio expansion solutions \cite{Freidberg1987,Fitzpatrick2024}.

Occupying the middle ground are inverse-coordinate moment solvers like VMOMS \cite{Lao1981, Lao1982, Lao1984} and the 3-moment approach \cite{Zakharov1986} used in transport codes like ASTRA \cite{Pereverzev2002}. These methods typically expand the poloidal flux surface geometry using standard Fourier series truncated to a few leading orders (e.g., up to $\sin(2\theta)$ or $\cos(3\theta)$) and solve the resulting one-dimensional (1D) radial ordinary differential equations (ODEs) via variational principles. While these approaches reduce the parameter count compared to two-dimensional (2D) grid-based solvers, their reliance on standard Fourier harmonics limits their compactness: efficiently representing the ``corners'' of a D-shaped plasma or an X-point still requires a prohibitive number of Fourier modes, which these low-order approximations fail to capture. Furthermore, the radial profiles in these codes are often resolved using grid-based finite differences, which further increases the number of required parameters.

Recent analytic tokamak equilibrium solutions \cite{Guazzotto2021} also adopt a semi-analytical approach but require the numerical solution of coefficients to fit the practical LCFS, often restricting the free functions $P$ and $F$ to be quadratic in $\psi$. With a typical CPU time of around 0.5s, these methods are not sufficiently fast or general enough for practical integrated modeling and real-time experimental fitting applications.

In this work, we address a fundamental question: \textit{What is the minimum number of free parameters required to accurately represent numerical solutions of the GS equation?}
We propose a unified spectral representation framework that bridges the gap between fast approximations and high-fidelity solutions. Instead of standard Fourier series or local grid-based methods, we employ the recent \textbf{Miller Extended Harmonic (MXH)} expansion \cite{Arbon2021} for the poloidal geometry. The MXH basis embeds the geometric topology (such as D-shape and triangularity) directly into the coordinate mapping via a harmonic distortion of the angle, rather than through additive superposition, which has been shown to achieve rapid convergence of shape parameters \cite{Snoep2023}. This is combined with Shifted Chebyshev polynomials for the radial profiles to ensure spectral accuracy and analytic smoothness. Consequently, both poloidal and radial representations converge rapidly with a minimal number of free parameters.

This approach offers a hierarchical structure:
\begin{itemize}
    \item \textbf{Approximation Level (System Code Mode):} At low truncation orders, our model naturally degenerates into the classic variational forms (similar to Haney), requiring only 2--5 parameters for a qualitative description.
    \item \textbf{High-Fidelity Level (Physics Analysis Mode):} By moderately increasing the resolution, the model converges exponentially to the accuracy of grid-based codes. We demonstrate that relative errors on the order of $10^{-2}$--$10^{-3}$ can be achieved with as few as 13--20 parameters, and complex equilibria can be fully resolved with fewer than 100 parameters.
\end{itemize}

The resulting equilibrium configurations are fully analytical, with smooth derivatives of all orders. These results provide a systematic foundation for developing high-fidelity, ultra-fast GS solvers and enable efficient reduced-order and AI-based surrogate modeling of tokamak equilibria.

The remainder of this paper is organized as follows. Section \ref{sec:basisfunc} establishes the theoretical framework, detailing the mathematical formulation of the Miller Extended Harmonic (MXH) parameterization and the Shifted Chebyshev radial basis. Section \ref{sec:hierarchy} outlines the hierarchical parameterization strategy, defining the parameter spaces for both the minimalist approximation (Level 1) and the high-fidelity representation (Level 2). In Section \ref{sec:numerical_results}, we present comprehensive numerical benchmarks against the standard grid-based code CHEASE, verifying the spectral convergence and accuracy for both standard and complex asymmetric equilibria. Section \ref{sec:discussion} discusses the broader implications of this compact representation for ultra-fast solvers and AI-based surrogate modeling. Finally, a summary and concluding remarks are provided in Section \ref{sec:summary}.

\section{Theoretical Framework and Basis Functions}\label{sec:basisfunc}

To address the challenge of minimizing the parameter space while maintaining high fidelity, we construct a spectral representation that decouples the poloidal geometry from the radial profiles. The formulation is established within the COCOS=1 convention \cite{Sauter2013} (counter-clockwise toroidal angle $\phi$, with the coordinate system $(R, \phi, Z)$ and $(\rho, \theta, \phi)$), where the toroidal plasma current $I_p$ is typically negative.

\begin{figure*}
\centering
\includegraphics[width=17.5cm]{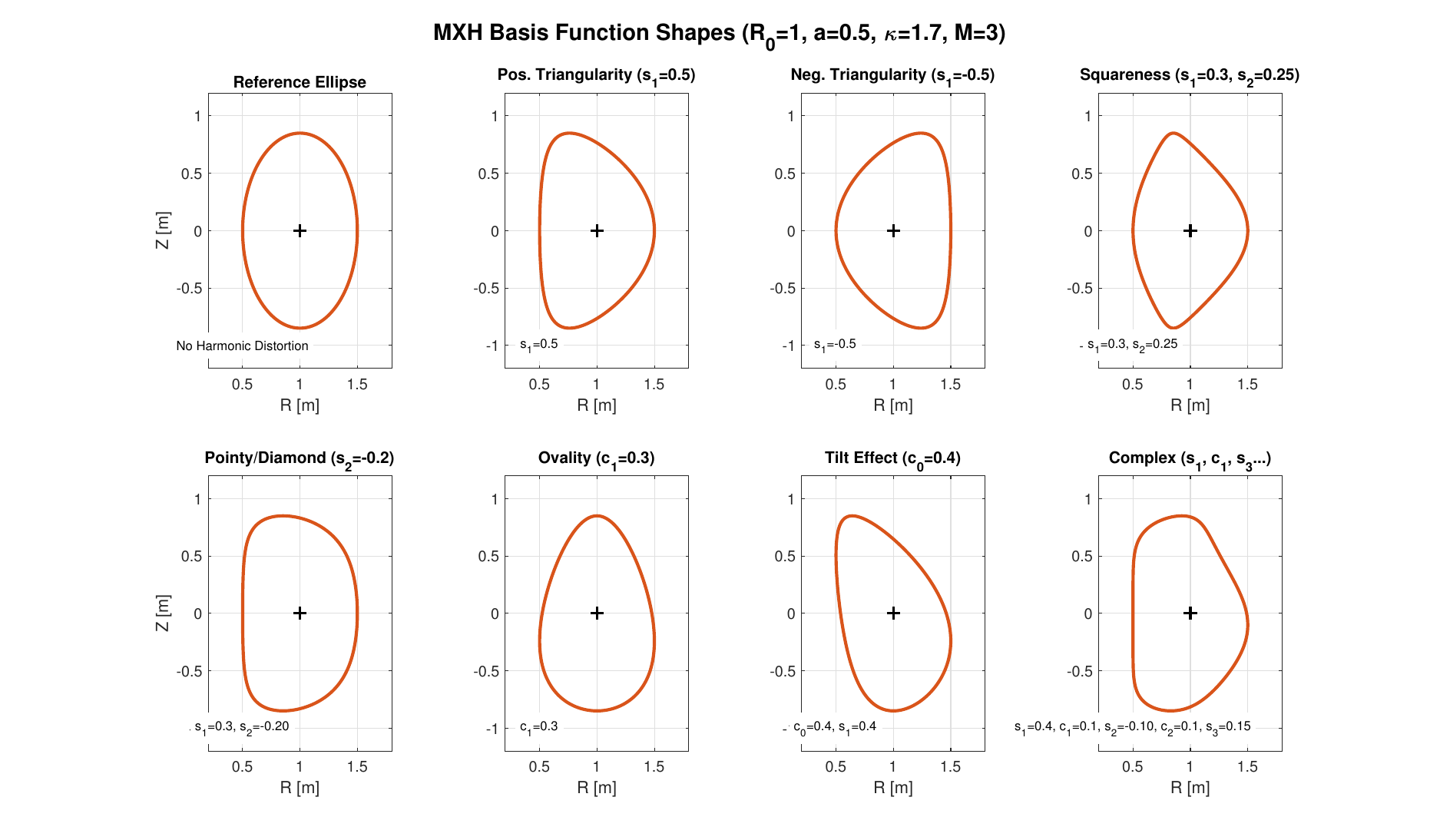}
\caption{Illustration of the Miller Extended Harmonic (MXH) poloidal parameterization. The panels demonstrate the topological flexibility of the basis by isolating specific harmonic coefficients. The \textbf{top row} displays standard symmetric shaping effects, including the reference ellipse, positive/negative triangularity ($\delta \approx \arcsin s_1$), and squareness ($s_2$). The \textbf{bottom row} illustrates asymmetric and higher-order deformations, including ovality ($c_1$), tilt ($c_0$), and complex combinations involving higher harmonics ($s_3$, etc.). This formulation allows for the analytic description of a wide range of tokamak geometries.}
\label{fig:mxh_shape}
\end{figure*}

\subsection{Radial Coordinate Definition and Mapping}
A critical aspect of the representation is the definition of the normalized radial coordinate $\rho \in [0, 1]$. The choice of $\rho$ determines the complexity of the profile functions and the form of the GS solver. We consider two primary definitions:

\begin{enumerate}
    \item \textbf{Flux-based Coordinate ($\rho_\psi$):} Defined as the square root of the normalized poloidal flux:
    \begin{equation}
    \label{eq:rho_psi}
    \rho_\psi \equiv \sqrt{\frac{\psi - \psi_{axis}}{\psi_{edge} - \psi_{axis}}}.
    \end{equation}
    In this frame, the flux function is analytically fixed as $\psi(\rho) = \psi_{axis} + (\psi_{edge}-\psi_{axis})\rho^2$. This simplifies the GS equation significantly as $\psi$ is known, but it requires the geometric minor radius $a(\rho)$ to be a free function to capture the flux surface shapes.
    
    \item \textbf{Geometric Coordinate ($\rho_{geom}$):} Defined based on the physical dimensions, typically $\rho_{geom} = r/a_{edge}$, where $r=(R_{max}-R_{min})/2$ is a characteristic minor radius. In this frame, the geometric profile $a(\rho)$ in Eq.~(\ref{eq:R_mxh}) becomes a constant ($a(\rho) = a_{edge}$), drastically reducing the number of radial parameters required to describe the geometry. However, this comes at the cost of making $\psi(\rho)$ an unknown function \cite{Haney1995} that must be solved for.
\end{enumerate}

The proposed framework supports both definitions. For fixed-boundary equilibrium reconstruction (fitting), $\rho_\psi$ is often preferred for its direct link to physical fluxes. For ultra-fast variational solvers, $\rho_{geom}$ can be advantageous due to the reduced geometric degrees of freedom.

\subsection{Poloidal Representation: MXH vs. Standard Fourier}\label{sec:basisfunc_mxh}
The efficient parameterization of the flux surface geometry $(R, Z)$ is the critical factor in reducing the number of free parameters. Standard spectral codes often employ a direct Fourier series expansion:
\begin{equation}
R(\theta) = \sum R_m \cos(m\theta), \quad Z(\theta) = \sum Z_m \sin(m\theta).
\end{equation}
While mathematically complete, this basis converges slowly for modern tokamak configurations characterized by high triangularity ($\delta$) and X-points. Capturing the sharp ``corners'' of a D-shaped plasma using smooth sinusoids typically induces the Gibbs phenomenon, requiring a large number of harmonics ($m \gg 10$) to suppress non-physical oscillations.

To overcome this, we employ the \textbf{Miller Extended Harmonic (MXH)} parameterization \cite{Arbon2021}. Instead of additive superposition, the MXH basis embeds the geometric topology directly into the coordinate mapping via an angular distortion. This basis has been further verified to perform well for flux-surface parameterization \cite{Snoep2023} and used in the TEQUILA 2D equilibrium model within the FUSE fusion power plant design code \cite{Meneghini2024}. The flux surface coordinates are parameterized as:
\begin{eqnarray}
R(\rho, \theta) &=& R_0 + h(\rho) + \rho \, a(\rho) \cos \bar{\theta}, \label{eq:R_mxh} \\
Z(\rho, \theta) &=& Z_0 + v(\rho) + \kappa(\rho) \, \rho \, a(\rho) \sin \theta, \label{eq:Z_mxh}
\end{eqnarray}
where $\theta \in [0, 2\pi)$ is the geometric poloidal angle. The geometric moments are functions of $\rho$: $h(\rho)$ is the Shafranov shift, $v(\rho)$ is the vertical displacement, $\kappa(\rho)$ is the elongation, and $a(\rho)$ describes the minor radius evolution. 

The key innovation lies in the distorted angle $\bar{\theta}$ used in the radial component, which is defined as:
\begin{equation}
\label{eq:theta_bar}
\bar{\theta} = \theta + c_0(\rho) + \sum_{m=1}^{M} \left[ c_m(\rho) \cos m\theta + s_m(\rho) \sin m\theta \right].
\end{equation}
This formulation naturally captures complex shaping with minimal terms:
\begin{itemize}
    \item The $m=1$ sine term ($s_1$) directly generates triangularity ($\delta \approx \arcsin s_1$). Unlike standard Fourier series which requires a cascade of harmonics to form a D-shape, MXH achieves this topologically with a single parameter.
    \item The $m=1$ cosine term ($c_1$) represents ovality.
    \item The $m=2$ terms relate to squareness ($s_2$) and higher-order deformations.
    \item The $c_0$ term represents the tilt of the configuration.
\end{itemize}
The coefficients $c_m$ and $s_m$ allow for the description of up--down asymmetric equilibria. By truncating the sum at small $M$ (e.g., $M=1$), the model reduces to the standard Miller shape \cite{Miller1998} used in fast system codes; increasing $M$ allows for spectral convergence to arbitrary shapes. Figure~\ref{fig:mxh_shape} illustrates typical LCFS shape fitting using the MXH basis.

\subsection{Radial Representation: Shifted Chebyshev Basis}
The geometric moments (e.g., $h(\rho), \kappa(\rho)$) and the physical source profiles ($P', FF'$) are expanded using Shifted Chebyshev polynomials of the first kind. This choice ensures optimal convergence properties and facilitates the handling of parity constraints.

Let $x$ be the mapped radial coordinate defined on the interval $[-1, 1]$:
\begin{equation}
x(\rho) = 2\rho^2 - 1, \quad {\rm for~} \rho \in [0, 1].
\end{equation}
We construct the radial basis functions $u_l(\rho)$ as:
\begin{equation}
\label{eq:basis_u}
u_l(\rho) = (1 - \rho^2) \cdot T_l(2\rho^2 - 1),
\end{equation}
where $T_l(x)$ is the Chebyshev polynomial of degree $l$, satisfying the recurrence relation $T_0(x) = 1$, $T_1(x) = x$, and $T_{l+1}(x) = 2x T_l(x) - T_{l-1}(x)$.
Any radial profile $f(\rho)$ is then expanded as:
\begin{equation}
\label{eq:radial_expansion}
f(\rho) = f_{edge} + \sum_{l=0}^{L} f_l u_l(\rho),
\end{equation}
where $f_{edge}$ is the value at the boundary (determined by the fixed LCFS) and $\{f_l\}$ are the free parameters to be optimized.

This choice of basis offers three distinct mathematical advantages:
\begin{enumerate}
    \item \textbf{Boundary Condition:} Since $u_l(\rho=1) = 0$ due to the $(1-\rho^2)$ factor, the condition $f(\rho=1) = f_{edge}$ is rigorously satisfied for any choice of spectral coefficients.
    \item \textbf{Regularity and Parity:} The basis functions depend on $\rho$ only through $\rho^2$. This ensures that all profiles are even functions of $\rho$, automatically satisfying the physical regularity condition $\partial f / \partial \rho |_{\rho=0} = 0$ at the magnetic axis without imposing additional constraints. This follows from the fact that $f=f(\psi)$ and $\psi |_{\rho\to0}\sim\rho^2$ near the magnetic axis, yielding $\frac{df}{d\rho} |_{\rho\to0}=\frac{df}{d\psi}\frac{d\psi}{d\rho} |_{\rho\to0}\sim\frac{df}{d\psi}(2\rho) |_{\rho\to0}=0$.
    \item \textbf{Minimax Property:} Chebyshev expansions minimize the maximum approximation error (minimax property) for smooth functions and allow for efficient computation via Fast Fourier Transform (FFT).
\end{enumerate}
A similar global radial basis is used in DESC (Fourier--Zernike basis) \cite{Dudt2020} to resolve the inaccuracy at the magnetic axis, as these global bases inherently satisfy the necessary conditions at the coordinate singularity.

\begin{figure*}
\centering
\includegraphics[width=17.5cm]{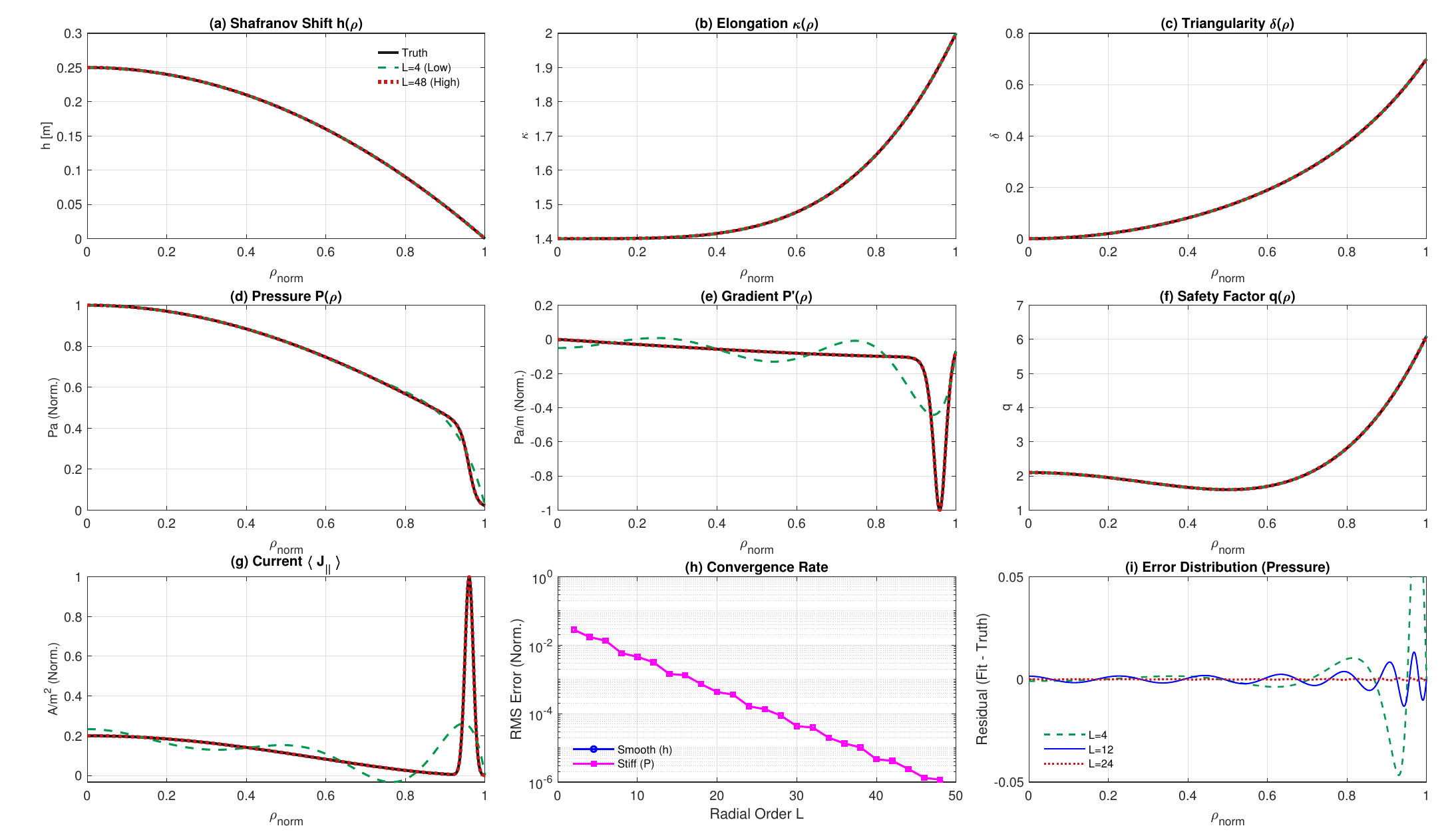}
\caption{Spectral convergence analysis of the Shifted Chebyshev radial basis. \textbf{(a--g)} Fitting of representative geometric profiles (e.g., Shafranov shift $h$, elongation $\kappa$) and stiff physical source profiles (e.g., Pressure $P$, Current $J_{\parallel}$) using low ($L=4$, green dashed) and high ($L=48$, red dotted) truncation orders compared to the ground truth (black solid). While smooth geometric moments are well-captured with low orders, stiff profiles with H-mode pedestals require higher orders to resolve gradients. \textbf{(h)} The root-mean-square (RMS) error versus radial order $L$ confirms exponential convergence for both smooth and stiff functions. \textbf{(i)} Residual distribution for the pressure profile, showing that increasing $L$ effectively minimizes Gibbs oscillations near the pedestal.}
\label{fig:cheb_fit}
\end{figure*}

Figure~\ref{fig:cheb_fit} demonstrates the typical Chebyshev fitting for radial profiles, exhibiting exponential convergence. For smooth profiles, $L \le 3$ is often sufficient. However, for strong gradients in $P$ or peaking in $J_{\parallel}$, $L \approx 20$--$50$ can accurately capture both the values and the derivatives, whereas grid-based solvers typically require $N_r \ge 256$ radial grid points. For the above choice of basis functions, the GS equation in inverse coordinates $(\rho, \theta)$ and relevant analytical derivatives are given in \ref{sec:app_gs_inv}.

\section{Hierarchical Parameterization Strategy}\label{sec:hierarchy}

The unified spectral framework defined in Section~\ref{sec:basisfunc} transforms the continuous equilibrium problem into a discrete parameter optimization problem. By adjusting the truncation orders of the poloidal harmonics ($M$) and the radial polynomials ($L$), we can systematically vary the complexity of the representation. This leads to a hierarchical strategy that seamlessly adapts to different accuracy requirements.

\subsection{The Parameter Space}
The total state of the tokamak equilibrium is uniquely determined by the spectral coefficients of the geometric profiles and, depending on the problem definition, the physical source profiles.
For a standard fixed-boundary equilibrium problem, the geometry of the last closed flux surface (LCFS) is specified, which fixes the boundary values ($f_{edge}$) for all geometric moments. The source functions---pressure gradient $P'(\psi)$ and poloidal current function $FF'(\psi)$---are typically provided as inputs. Consequently, the free parameters are primarily the coefficients of the internal flux surface geometry.

We define the global parameter vector $\mathbf{X}$ as the concatenation of coefficients for:
\begin{itemize}
    \item \textbf{Geometric profiles:} The radial variations of the Shafranov shift $h(\rho)$, vertical position $v(\rho)$, elongation $\kappa(\rho)$, and the shaping harmonics $\{c_m(\rho), s_m(\rho)\}_{m=1}^M$. 
    \item \textbf{Radial Coordinate Map:} Either the minor radius profile $a(\rho)$ (if using the flux-based coordinate $\rho_\psi$) or the flux profile $\psi(\rho)$ (if using the geometric coordinate $\rho_{geom}$).
    \item \textbf{Scalar unknowns:} The total flux scalar $\psi_a \equiv \psi_{edge} - \psi_{axis}$.
\end{itemize}
The dimension of $\mathbf{X}$, denoted as $N_{par}$, is the fundamental quantity we seek to minimize.

\subsection{Level 1: The Minimalist Representation (Approximation)}
This level targets the regime of ``System Codes'' \cite{Haney1992,Chen2026} or real-time control, where computational speed is paramount and a qualitative description of the core plasma suffices.

\begin{itemize}
    \item \textbf{Settings:} We truncate the poloidal expansion at $M=1$ and the radial expansion at low order (typically $L=1$ or $2$).
    \item \textbf{Physics:} Retaining only $M=1$ restricts the geometry to a Miller-like shape (characterized by shift, elongation, and triangularity). By limiting radial variations to $L=1$--$2$, the profiles are constrained to be parabolic or monotonic, capturing the global moments but smoothing over local gradients.
    \item \textbf{Equivalence:} In this regime, our representation becomes mathematically equivalent to the variational moment models described by Haney \cite{Haney1988}, or the more complex formulations by Lao \cite{Lao1981} and Zakharov \cite{Zakharov1986}.
    \item \textbf{Parameter Count:} The number of free parameters can be as low as $N_{par} \approx 3$--$5$. For instance, if profiles are assumed to be stiff, only the core values of shift and elongation are required alongside the flux scalar.
\end{itemize}

At this level, the proposed spectral solver offers superior accuracy compared to the analytical moment solutions used in fast integrated modeling tools like METIS \cite{Artaud2018}, which often rely on large-aspect-ratio expansions \cite{Lao1981}. Furthermore, by allowing for flexible radial basis orders ($L$), this method generalizes the 3-moment approach \cite{Zakharov1986} used in ASTRA, enabling a more accurate description of the Shafranov shift and elongation profiles without the need for solving grid-based ODEs.

\subsection{Level 2: The High-Fidelity Representation (Complete)}
This level targets the regime of ``Physics Analysis,'' such as MHD stability (e.g., peeling-ballooning modes) or transport modeling, where the precise curvature of flux surfaces and local gradients are critical.

\begin{itemize}
    \item \textbf{Settings:} We increase the poloidal resolution to $M \approx 3$--$6$ and radial resolution to $L \approx 6$--$12$.
    \item \textbf{Physics:}
        \begin{enumerate}
            \item \textbf{Poloidal:} Higher harmonics ($m \ge 2$) in the MXH angle $\bar{\theta}$ capture ``squareness'' ($s_2$), ovality, and the local deformations induced by X-points. As discussed in Section~\ref{sec:basisfunc_mxh}, the MXH basis converges significantly faster than standard Fourier series here; $M=4$ in MXH often matches the fidelity of $m \sim 20$ in standard Fourier representations.
            \item \textbf{Radial:} Higher-order Chebyshev polynomials allow for the description of internal transport barriers (ITBs) or steep edge pedestals with high fidelity.
        \end{enumerate}
    \item \textbf{Parameter Count:} We find that $N_{par} \approx 13$--$20$ is sufficient to reproduce the results of grid-based solvers (such as CHEASE) for standard D-shaped plasmas with relative errors on the order of $10^{-3}$. We also observe that reference solutions near the magnetic axis and edge in grid-based codes may be inaccurate due to the violation of physical analyticity features, such as the triangularity condition $ds_1/d\rho|_{\rho\to0} \to 0$, which our spectral basis naturally enforces.
\end{itemize}

For higher accuracy requirements, such as resolving steep edge gradients or complex X-point geometries, $N_{par}$ may increase to $80$--$100$. However, this is still two orders of magnitude less than the degrees of freedom required by grid-based solvers such as CHEASE and ECOM.

\subsection{Optimization Goal}
The problem of finding the ``minimum number of parameters'' is thus formulated as finding the smallest integers $M$ and $L$ (and consequently the smallest vector $\mathbf{X}$) that approximate the solution of the GS equation within a given tolerance. The definition of ``best approximation'' depends on the solver strategy:
\begin{enumerate}
    \item \textbf{Variational Method:} Minimizing the Lagrangian of the MHD energy functional, similar to VMOMS \cite{Lao1981} or Haney \cite{Haney1995}.
    \item \textbf{Residual Method:} Minimizing the $L_2$ norm of the GS equation residual, similar to DESC \cite{Dudt2020}.
\end{enumerate}
In practical tests, for small $N_{par} \le 10$, we find the variational method performs better; for example, it can capture the Shafranov shift of the magnetic axis more accurately than the residual method. For large $N_{par}$, the residual method is generally preferred \cite{Dudt2020}.

For the residual minimization approach, the optimization problem is stated as:
\begin{equation}
    \min_{\mathbf{X}} \dim(\mathbf{X}) \quad {\rm subject~ to~ } ||\Delta^* \psi(\mathbf{X}) + \mu_0 R J_\phi(\mathbf{X})|| < \epsilon,
\end{equation}
where $\epsilon$ is a specified tolerance. In the following section, we demonstrate numerical solutions to this problem, benchmarking the results against standard high-fidelity codes.

\begin{figure*}
\centering
\includegraphics[width=17.5cm]{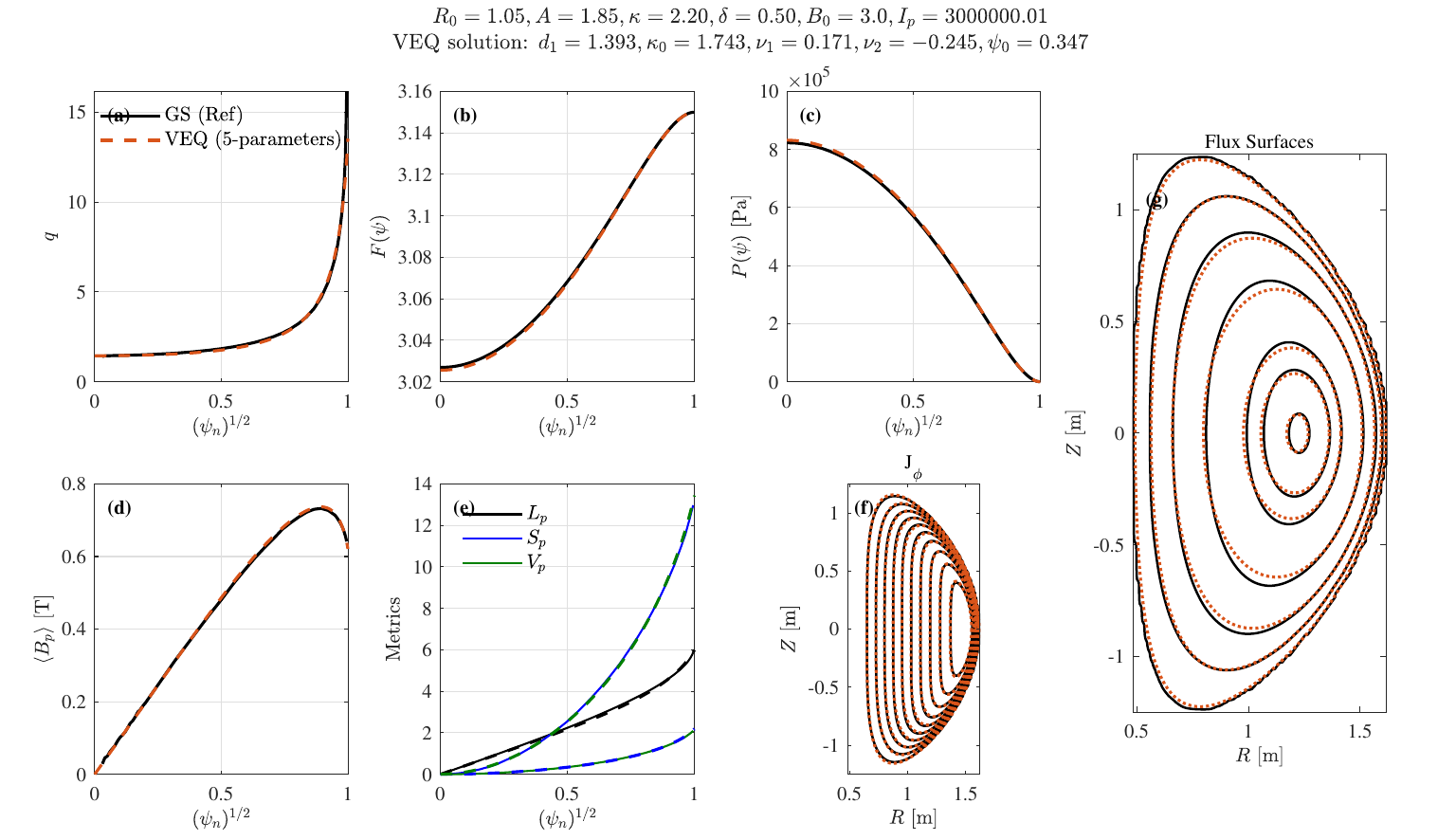}
\caption{Level 1 Validation: Comparison between the reduced-order variational solver VEQ (Red Dashed) and a full grid-based GS solver (Black Solid). The VEQ model uses a minimal set of only \textbf{5 free parameters} ($d_1, \kappa_0, \nu_1, \nu_2, \psi_a$) to approximate the equilibrium. Despite the extreme compression, the 1D physical profiles---including safety factor $q$, poloidal field $B_p$, plasma volume $V_p$, surface area $S_p$, and poloidal arc length $L_p$---show excellent agreement (relative error $< 5\%$) with the high-fidelity numerical solution, validating its suitability for system codes.}
\label{fig:veq_gs_cmp}
\end{figure*}

\begin{figure*}
\centering
\includegraphics[width=17.5cm]{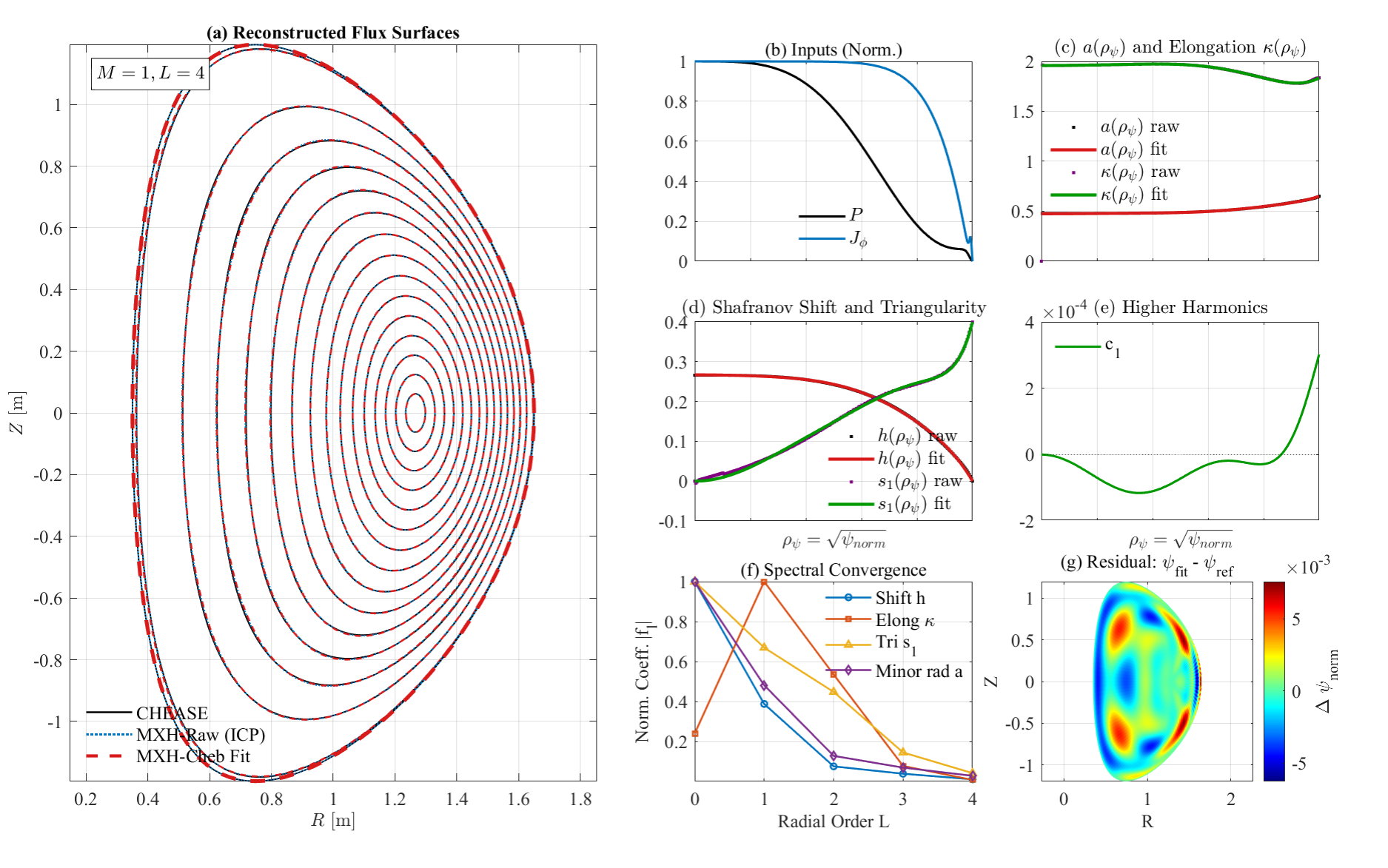}
\caption{Level 2 Validation (Standard Configuration): Spectral representation of a high-fidelity CHEASE equilibrium characterized by a standard up-down symmetric D-shape. (a) The reconstructed 2D flux surfaces (Red Dashed) visually overlap perfectly with the reference solution (Black Solid). (b-g) Radial profiles of the geometric moments and input physics. For this standard configuration, the equilibrium is accurately represented ($\epsilon \sim 10^{-3}$) using a truncated basis with $M=1$ (poloidal) and $L \le 4$ (radial), requiring only $\approx \textbf{13--16 parameters}$.}
\label{fig:chease_dat1}
\end{figure*}

\begin{figure*}
\centering
\includegraphics[width=17.5cm]{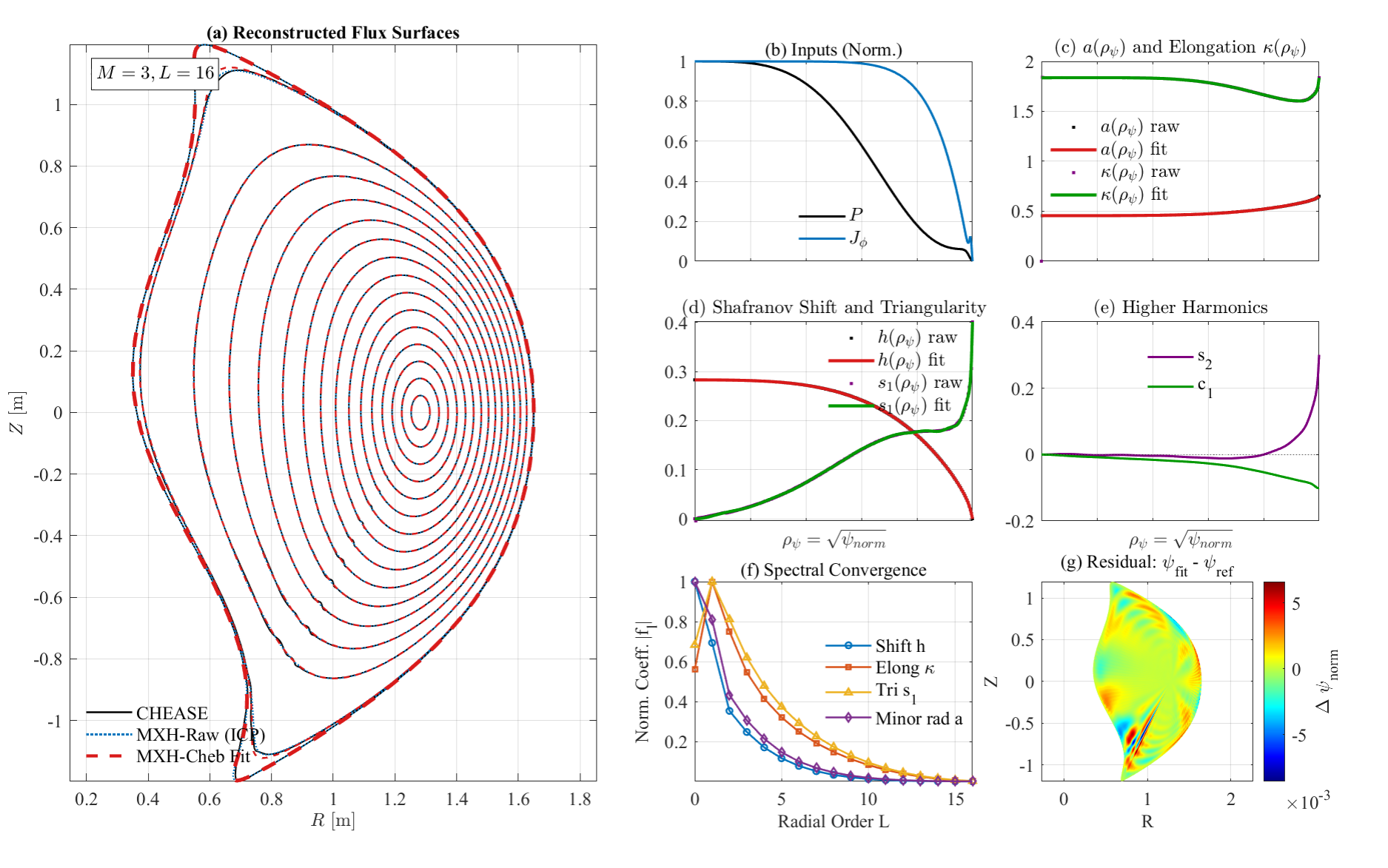}
\caption{Level 2 Validation (Complex Asymmetric Configuration): High-fidelity representation of a single-null divertor equilibrium featuring significant up-down asymmetry. (a) 2D flux surface reconstruction using the Iterative Closest Point (ICP) refinement for boundary matching. (g) The "Higher Harmonics" panel reveals non-zero radial profiles for squareness ($s_2$) and ovality ($c_1$), confirming that simple $M=1$ models are insufficient for this regime. Achieving this level of fidelity (resolving the X-point vicinity and stiff edge gradients) typically requires extending the basis to $M \approx 3$ and $L \approx 16$, utilizing approximately $\mathbf{85}$ parameters.}
\label{fig:chease_dat2}
\end{figure*}

\section{Numerical Results and Validation} \label{sec:numerical_results}

To quantify the efficiency and accuracy of the proposed representation, we benchmark the MXH-Cheb basis against high-fidelity equilibria generated by the fixed-boundary code CHEASE \cite{Lutjens1996}. The validation is performed at two hierarchical levels corresponding to the strategy outlined in Section \ref{sec:hierarchy}. 

It is important to distinguish between two different objectives in this section:
\begin{itemize}
    \item \textbf{Forward Solver (Level 1):} Solving the variational problem to find the equilibrium state from scratch using a minimal parameter set. This tests the physical validity of the reduced model.
    \item \textbf{Inverse Representation (Level 2):} Fitting an existing high-fidelity numerical solution (CHEASE) to find the most compact spectral coefficients. This tests the mathematical completeness and convergence of the basis functions.
\end{itemize}

\subsection{Level 1: Variational Parameter GS Solver}

At the approximation level, we restrict the representation to the lowest order moments to emulate the requirements of fast system codes. We adopt a 3-moment geometric approximation where the radial coordinate is defined geometrically as $\rho=\rho_{geom}$ with a constant minor radius $a(\rho)=a$.

To minimize the degrees of freedom, the geometric moments are approximated using low-order polynomials:
\begin{eqnarray}
h(\rho) &=& d_1(1-\rho^2), \\
\kappa(\rho) &=& \kappa_0+(\kappa_a-\kappa_0)\rho^4, \\
s_1(\rho) &=& s_{1a}\rho^2, \\
\psi(\rho) &=& \psi_a[\nu_1\rho^2+\nu_2\rho^4+(1-\nu_1-\nu_2)\rho^6],
\end{eqnarray}
where the boundary parameters $(\kappa_a, s_{1a})$ are inputs defining the LCFS. This formulation results in only 5 free parameters to be solved: $(d_1, \kappa_0, \nu_1, \nu_2, \psi_a)$. The primary difference between our model and Haney's model \cite{Haney1988,Haney1995} is twofold: we remove the free parameters for triangularity profiles as they are less sensitive to the variational energy, and we add a higher-order parameter $\nu_2$ to the flux function $\psi(\rho)$ to more accurately capture the poloidal magnetic field $B_p$ and safety factor $q(\rho)$. Furthermore, explicit high-order configuration functions for up-down asymmetry and X-points used in \cite{Haney1995} are not necessary here, as the general MXH basis can provide these shapes naturally if the model is extended.

We developed the VEQ (`V' stands for Variational or Veloce) code to solve these 5 parameters by minimizing the MHD Lagrangian. Figure \ref{fig:veq_gs_cmp} presents a typical comparison with a grid-based GS solver. Despite the extreme compression to 5 parameters, the agreement is remarkable. Although the 2D shape of $\psi(R,Z)$ shows slight deviations from the accurate numerical solution, the 1D physics profiles--such as the safety factor $q(\rho)$, poloidal magnetic field $B_p(\rho)$, and macroscopic quantities ($V_p, S_p, L_p$)--agree well. The magnetic axis position is also captured accurately. The overall disagreement is less than 5\% for most test cases. The typical CPU time to solve these parameters in VEQ is less than 50 ms \cite{Chen2026}, with further optimization possible.

\subsection{Level 2: High-Fidelity Representation}

For applications requiring precise local curvature and gradient information (e.g., stability analysis), we utilize the full MXH-Cheb basis to represent high-fidelity equilibria produced by CHEASE.

\subsubsection{Geometric Fitting Algorithm (ICP)}
A critical challenge in benchmarking against grid-based codes is the definition of the geometric angle. Initial attempts using the direct bounding-box fitting workflow proposed in the original MXH work \cite{Arbon2021} yielded geometric residuals only on the order of $10^{-2}$, particularly for strongly shaped plasmas. To overcome this limit, we implemented an Iterative Closest Point (ICP) refinement algorithm. By iteratively optimizing the geometric parameters to minimize the Euclidean distance between the MXH curve and the discrete data points (rather than relying on fixed angular parameterization), we reduced the geometric fitting error to $\sim 5 \times 10^{-4}$, enabling a nearly exact match of the LCFS. This step is essential for establishing a high-fidelity ground truth for the spectral representation.

\subsubsection{Fitting Standard Equilibria (3-Moment LCFS)}
Standard system codes often utilize a 3-moment description of the boundary. We demonstrate that for such configurations, the internal radial profiles can be accurately represented by as few as 13--16 parameters. This corresponds to truncating the basis at $M=1$ (poloidal) and $L \le 4$ (radial).

Figure \ref{fig:chease_dat1} illustrates the fitting results for a standard D-shaped plasma ($\kappa \approx 1.7$, $\delta \approx 0.4$). The spectral fit (Red Dashed) visually overlaps perfectly with the CHEASE reference (Black Solid). The radial profiles of geometric moments are smooth and capture the core variation accurately.

\subsubsection{Fitting Complex Asymmetric Equilibria}
For experimental configurations involving single-null divertors, the plasma shape becomes up-down asymmetric and possesses an X-point. This requires higher-order harmonics.

Figure \ref{fig:chease_dat2} shows the fitting results for a complicated, asymmetric LCFS. To capture the "squareness" and the local deformation near the X-point, we increase the poloidal resolution to $M=3$ and radial resolution to $L=16$. 
The "Higher Harmonics" panel in Fig.~\ref{fig:chease_dat2} reveals that while the amplitudes of higher-order MXH harmonics ($s_2, c_1, \dots$) are small compared to the primary moments, they are non-zero and essential for achieving the $10^{-3}$ accuracy level. Specifically, $c_1$ captures the up-down asymmetry (ovality), and $s_2$ captures the squareness.
Regarding the radial profiles, to handle strong gradients in the input pressure $P(\rho)$ and toroidal current density $J_{\phi}(\rho)$ (e.g., H-mode pedestals), higher radial orders ($L \sim 12-16$) are required. Consequently, a typical high-fidelity representation for a complex tokamak equilibrium requires $N_{par} \approx 85$ parameters.

\subsection{Convergence and Comparison with Standard Methods}

We evaluated the relative error $\epsilon = ||\psi_{fit} - \psi_{ref}|| / ||\psi_{ref}||$ as a function of the total number of free parameters $N_{par}$.

\textbf{Analysis of Residual Sources:} 
While the geometric fitting of the LCFS achieves errors as low as $5 \times 10^{-4}$, the global reconstruction residual $\epsilon$ against CHEASE data typically saturates around $5 \times 10^{-3}$. This discrepancy is not solely due to the basis truncation but involves data quality and physical consistency issues. First, visual inspection of the dense CHEASE grid ($N_\psi \times N_\theta = 257 \times 1025$) reveals numerical oscillations near the X-point region ($\theta \sim 2\pi$), suggesting inaccuracies in the reference solution itself. Second, the raw geometric parameters extracted from CHEASE often exhibit non-zero radial derivatives at the magnetic axis (e.g., $d\delta/d\rho|_{\rho=0} \neq 0$), violating the physical analyticity condition. Our spectral representation enforces the correct parity constraints (zero derivatives at the axis), naturally leading to deviations from the "noisy" or non-analytic reference data. Thus, the spectral representation acts as a physics-informed filter, correcting the numerical artifacts of the grid-based solution.

\textbf{Comparison with Fourier Series:} Standard Fourier representations typically require $m > 12$ harmonics to resolve a D-shape without significant Gibbs ringing. In contrast, the MXH-Cheb method achieves comparable fidelity with $N_{par} < 100$. This represents a compression ratio of nearly two orders of magnitude compared to grid-based methods ($10^4$ DOF).

In summary, the "minimum number of parameters" depends on the application:
\begin{itemize}
    \item \textbf{Min $\approx$ 3--5:} For qualitative physics and global scoping (System Codes).
    \item \textbf{Min $\approx$ 15--20:} For quantitative analysis indistinguishable from standard grid codes for most experimental purposes.
    \item \textbf{Min $\approx$ 80--100:} For fully resolving complex X-point geometries and stiff H-mode profiles.
\end{itemize}

\section{Discussion: Implications for Ultra-fast GS Solvers and AI}\label{sec:discussion}

The results presented in Section~\ref{sec:numerical_results} have profound implications beyond standard equilibrium analysis, particularly for the emerging fields of data-driven plasma science and real-time control.

\subsection{Overcoming the Curse of Dimensionality for AI}
In the context of machine learning (ML) for fusion, a major bottleneck is the high dimensionality of the target data. Training a neural network to predict a 2D flux map $\psi(R,Z)$ represented on a $65 \times 65$ grid involves an output layer of over 4000 neurons. This ``curse of dimensionality'' often requires massive training datasets to avoid overfitting and makes the training process computationally expensive.

By employing the MXH-Cheb representation, we effectively perform a physics-informed compression. Instead of learning the grid values, a neural network needs only to learn the mapping from engineering actuators (currents, field coils) to the small parameter vector $\mathbf{X}$ (size $\sim 15$--$100$). This reduction by two orders of magnitude significantly lowers the data requirements for training reliable surrogate models.

\subsection{Efficient Database Storage}
Fusion experimental databases and integrated modeling codes often store thousands of equilibrium time-slices, imposing a heavy storage burden when represented as full 2D grids. The proposed Level 2 representation serves as a ``lossless'' compression format (physically indistinguishable from the grid data within the error margin), allowing for the storage of entire discharge campaigns in a fraction of the space. Moreover, unlike grid data, this format retains the analytic derivatives, enabling precise reconstruction of magnetic fields $(B_R, B_Z)$ and current densities at any spatial point without interpolation errors.

\subsection{Toward Ultra-Fast Direct Solvers}
While this work focuses on \textit{representation} (inverse problem: fitting $\mathbf{X}$ given $\psi$), the superior convergence properties of the MXH-Cheb basis suggest it is also the ideal candidate for a direct \textit{solver} (forward problem). A spectral solver based on this basis (using Galerkin or collocation methods) could potentially solve the GS equation in milliseconds, offering a physics-based alternative to black-box neural networks for real-time applications. By leveraging modern numerical frameworks such as GPU acceleration and automatic differentiation (e.g., JAX), the solution time could possibly reduce to the sub-millisecond regime. The detailed implementation of such a solver is a subject of future work. This approach differs from DESC \cite{Dudt2020} by utilizing the more rapidly convergent poloidal MXH basis specifically optimized for tokamak shapes.

\subsection{Open Questions}
An open question remains: Can we find even better basis functions than MXH-Cheb to further reduce the free parameter requirement? While MXH captures D-shapes and squareness efficiently, extremly complex geometries (e.g., snowflakes or negative triangularity with sharp corners) might benefit from domain decomposition or adaptive spectral element methods to avoid high-order global polynomials.

\section{Summary and Conclusion}\label{sec:summary}

In this work, we addressed the fundamental question: \textit{What is the minimum number of free parameters required to represent solutions of the Grad-Shafranov equation?}

By introducing a unified spectral framework combining the Miller Extended Harmonic (MXH) parameterization for poloidal geometry and Shifted Chebyshev polynomials for radial profiles, we demonstrated that the complexity of tokamak equilibria is far lower than traditional grid-based methods suggest. Our findings are summarized as follows:

\begin{enumerate}
    \item \textbf{Hierarchical Structure:} The method provides a continuous transition from low-order approximations to high-fidelity solutions within a single mathematical framework.
    \item \textbf{Level 1 (System Code Regime):} For global scoping studies, the equilibrium can be adequately described by \textbf{3--5 parameters} (essentially shift, elongation, triangularity, and profile scalars). This validates and generalizes the physical intuition behind early variational models like Haney et al.
    \item \textbf{Level 2 (High-Fidelity Regime):} For precision physics (stability/transport), relative errors of $10^{-2}$--$10^{-3}$ are achievable with approximately \textbf{15--20 parameters} for standard shapes. For complex asymmetric configurations with stiff profiles, $\sim 80$--$100$ parameters are sufficient to capture details such as X-points and H-mode pedestals. This is in stark contrast to standard Fourier/grid methods which require $10^3 \sim 10^5$ degrees of freedom.
    \item \textbf{Analyticity and Physics Compliance:} Unlike piecewise-linear grid solutions, the reconstructed equilibria are fully analytic and infinitely differentiable. Our analysis reveals that the spectral representation acts as a filter, enforcing physical constraints (such as zero derivatives at the magnetic axis) that are often violated by numerical noise in grid-based codes.
\end{enumerate}

We conclude that for the vast majority of practical applications in tokamak design and modeling, the ``minimum number of parameters'' lies between 5 and 20. This compact representation opens new avenues for ultra-fast equilibrium solvers and highly efficient AI-based surrogate models \cite{Zheng2025}, facilitating the optimization of future fusion reactors.

\ack
The authors are grateful to Wei-qi Meng for providing the CHEASE data. Y. Y. Li acknowledges support from the National Natural Science Foundation of China (Grant No. 12405265). The authors also acknowledge the use of AI tools for assisting with the organization and language refinement of this manuscript.

\appendix
\section{Grad-Shafranov Equation in Inverse Coordinates}\label{sec:app_gs_inv}

In this appendix, we provide the explicit forms of the Grad-Shafranov (GS) equation transformed into the computational $(\rho, \theta)$ domain, along with the analytic derivatives of the MXH basis functions required to construct the metric tensor.

\subsection{Metric Coefficients and Jacobian}
Let the mapping from flux coordinates to physical coordinates be $R = R(\rho, \theta)$ and $Z = Z(\rho, \theta)$. We define the Jacobian of the transformation $J$ (oriented to be positive) as:
\begin{equation}
J = R_\rho Z_\theta - R_\theta Z_\rho,
\end{equation}
where subscripts denote partial derivatives (e.g., $R_\rho \equiv \partial R / \partial \rho$). Note that the sign convention here ensures $J>0$, assuming a standard right-handed poloidal angle definition.

The covariant metric coefficients in the poloidal plane are:
\begin{eqnarray}
g_{\rho\rho} &=& R_\rho^2 + Z_\rho^2, \\
g_{\theta\theta} &=& R_\theta^2 + Z_\theta^2, \\
g_{\rho\theta} &=& R_\rho R_\theta + Z_\rho Z_\theta.
\end{eqnarray}
Consequently, the gradient operators in the inverse coordinates are given by:
\begin{equation}
|\nabla \rho|^2 = \frac{g_{\theta\theta}}{J^2}, \quad |\nabla \theta|^2 = \frac{g_{\rho\rho}}{J^2}, \quad \nabla \rho \cdot \nabla \theta = -\frac{g_{\rho\theta}}{J^2}.
\end{equation}

\subsection{Transformed Equilibrium Equation}
The GS operator $\Delta^* \psi = R^2 \nabla \cdot (R^{-2} \nabla \psi)$ is transformed into the $(\rho, \theta)$ coordinates. Assuming the flux function depends only on $\rho$ (i.e., $\psi = \psi(\rho)$), the operator simplifies to:
\begin{equation}
\Delta^* \psi = \frac{R}{J} \left[ \frac{\partial}{\partial \rho} \left( \psi_\rho \frac{g_{\theta\theta}}{JR} \right) - \frac{\partial}{\partial \theta} \left( \psi_\rho \frac{g_{\rho\theta}}{JR} \right) \right].
\end{equation}
Substituting this into the equilibrium condition, we obtain the residual equation $G$ that must be minimized by the spectral coefficients:
\begin{eqnarray}
G &\equiv& \frac{R}{J} \left[ \frac{\partial}{\partial \rho} \left( \psi_\rho \frac{R_\theta^2 + Z_\theta^2}{JR} \right) - \frac{\partial}{\partial \theta} \left( \psi_\rho \frac{R_\rho R_\theta + Z_\rho Z_\theta}{JR} \right) \right] \nonumber \\
&& + \frac{1}{\psi_\rho} (FF' + R^2 \mu_0 P') = 0.
\end{eqnarray}
For the specific mapping $\psi(\rho) = \psi_a \rho^2$ used in this work, the derivative is simply $\psi_\rho = 2\psi_a \rho$.

\subsection{Derivatives of the MXH Geometry}
To evaluate the metrics, we require the derivatives of the coordinate functions. The MXH parameterization is given by:
\begin{eqnarray}
R(\rho, \theta) &=& R_0 + h(\rho) + A(\rho) \cos \bar{\theta}, \\
Z(\rho, \theta) &=& Z_0 + v(\rho) + K(\rho) \sin \theta,
\end{eqnarray}
where we define $A(\rho) \equiv \rho \, a(\rho)$ and $K(\rho) \equiv \rho \, a(\rho) \, \kappa(\rho)$ for compactness. The distorted angle is $\bar{\theta} = \theta + c_0(\rho) + \sum_{m=1}^M (c_m \cos m\theta + s_m \sin m\theta)$.

\subsubsection{First-Order Derivatives}
The derivatives with respect to the computational coordinates are:
\begin{eqnarray}
R_\theta &=& -A(\rho) \sin \bar{\theta} \cdot \bar{\theta}_\theta, \\
Z_\theta &=& K(\rho) \cos \theta, \\
R_\rho &=& h'(\rho) + A'(\rho) \cos \bar{\theta} - A(\rho) \sin \bar{\theta} \cdot \bar{\theta}_\rho, \\
Z_\rho &=& v'(\rho) + K'(\rho) \sin \theta.
\end{eqnarray}
Here, the derivatives of the angle function are:
\begin{eqnarray}
\bar{\theta}_\theta &=& 1 + \sum_{m=1}^M m(-c_m \sin m\theta + s_m \cos m\theta), \\
\bar{\theta}_\rho &=& c_0'(\rho) + \sum_{m=1}^M (c_m' \cos m\theta + s_m' \sin m\theta).
\end{eqnarray}

\subsubsection{Second-Order Derivatives}
These terms are essential for constructing the Jacobian of the residual vector in Newton-Krylov solvers or for computing magnetic curvature.

For the $Z$-coordinate:
\begin{eqnarray}
Z_{\theta\theta} &=& -K(\rho) \sin \theta, \\
Z_{\rho\theta} &=& K'(\rho) \cos \theta, \\
Z_{\rho\rho} &=& v''(\rho) + K''(\rho) \sin \theta.
\end{eqnarray}

For the $R$-coordinate:
\begin{eqnarray}
R_{\theta\theta} &=& -A \left[ \cos \bar{\theta} (\bar{\theta}_\theta)^2 + \sin \bar{\theta} \, \bar{\theta}_{\theta\theta} \right], \\
R_{\rho\theta} &=& -A' \sin \bar{\theta} \, \bar{\theta}_\theta - A \left[ \cos \bar{\theta} \, \bar{\theta}_\rho \bar{\theta}_\theta + \sin \bar{\theta} \, \bar{\theta}_{\rho\theta} \right], \\
R_{\rho\rho} &=& h'' + A'' \cos \bar{\theta} - 2A' \sin \bar{\theta} \, \bar{\theta}_\rho \nonumber \\
&& - A \left[ \cos \bar{\theta} (\bar{\theta}_\rho)^2 + \sin \bar{\theta} \, \bar{\theta}_{\rho\rho} \right].
\end{eqnarray}
Note that $\bar{\theta}_{\theta\theta}$, $\bar{\theta}_{\rho\theta}$, and $\bar{\theta}_{\rho\rho}$ are obtained by direct differentiation of the harmonic series.

\subsection{Radial Basis Derivatives}
The profile functions (e.g., $h(\rho), A(\rho)$) are expanded as $f(\rho) = \sum_{l=0}^L f_l u_l(\rho)$. The derivatives with respect to $\rho$ are computed via the recurrence relations of the Shifted Chebyshev polynomials.

Let $x = 2\rho^2 - 1 \in [-1, 1]$. The basis functions are $u_l(\rho) = (1-\rho^2) T_l(x)$. The derivatives follow from the chain rule:

\noindent \textbf{First derivative ($f_\rho$):}
\begin{equation}
\frac{d u_l}{d \rho} = -2\rho T_l(x) + 4\rho(1-\rho^2) \frac{dT_l}{dx}.
\end{equation}

\noindent \textbf{Second derivative ($f_{\rho\rho}$):}
\begin{equation}
\frac{d^2 u_l}{d \rho^2} = -2 T_l(x) + (4 - 20\rho^2) \frac{dT_l}{dx} + 16\rho^2(1-\rho^2) \frac{d^2 T_l}{dx^2}.
\end{equation}

The derivatives of the Chebyshev polynomials, denoted $T'_{l}(x) \equiv dT_l/dx$ and $T''_{l}(x) \equiv d^2T_l/dx^2$, are generated recursively:
\begin{eqnarray}
T'_{l}(x) &=& 2 T_{l-1}(x) + 2x T'_{l-1}(x) - T'_{l-2}(x), \\
T''_{l}(x) &=& 4 T'_{l-1}(x) + 2x T''_{l-1}(x) - T''_{l-2}(x).
\end{eqnarray}
Initialization values are:
\begin{itemize}
    \item $l=0$: $T_0=1, T'_{0}=0, T''_{0}=0$.
    \item $l=1$: $T_1=x, T'_{1}=1, T''_{1}=0$.
\end{itemize}
This recursive formulation allows for the efficient and exact evaluation of radial derivatives to arbitrary order. Note also that the derivatives to $\{f_l\}$ can also be obtained analytically, which could be helpful to numerically solve the values of the parameters in ${\bf X}$.

\end{document}